\documentclass{article}

\PassOptionsToPackage{numbers, compress}{natbib}

\usepackage[final]{neurips_2019}

\usepackage[utf8]{inputenc} 
\usepackage[T1]{fontenc}    
\usepackage{hyperref}       
\usepackage{url}            
\usepackage{booktabs}       
\usepackage{amsfonts}       
\usepackage{nicefrac}       
\usepackage{microtype}      

\usepackage{algorithm}      
\usepackage{algorithmic}    
\usepackage{graphicx}       
\usepackage{xcolor}         

\usepackage{subfig}
\usepackage{xspace}
\usepackage{amsmath}
\usepackage{comment}
\usepackage{enumitem}
\usepackage{times}
\usepackage{wrapfig}
\usepackage[subtle]{savetrees}

\newcommand{\todo}[1]{\textcolor{red}{todo: #1}}
\newcommand{\edit}[1]{\textcolor{black}{#1}}
\newcommand{\tim}[1]{\textcolor{blue}{tim: #1}}
\newcommand{\sanchit}[1]{\textcolor{purple}{sanchit: #1}}
\newcommand{\jialin}[1]{\textcolor{orange}{jialin: #1}}
\newcommand{\vikram}[1]{\textcolor{brown}{vikram: #1}}

\newcommand{\GENIE}{LISA\xspace}
\newcommand{\Figure}[1]{Fig.~\ref{fig:#1}}
\newcommand{\Table}[1]{Tab.~\ref{tab:#1}}
\newcommand{\Algorithm}[1]{Alg.~\ref{alg:#1}}
\newcommand{\Section}[1]{\S\ref{sec:#1}}

\usepackage{xspace}

\newcommand{\partCPUtm}{Intel\textsuperscript{\textregistered} Core\textsuperscript{\texttrademark}\xspace}

\title{\GENIE: Towards Learned DNA Sequence Search}

\author{%
    Darryl Ho \\
    MIT\\
    \texttt{\small darry140@mit.edu} \\
    \And
    Jialin Ding \\
    MIT \\
    \texttt{\small jialind@mit.edu} \\
    \And
    Sanchit Misra \\
    Intel Labs \\
    \texttt{\small sanchit.misra@intel.com} \\
    \And
    Nesime Tatbul \\
    Intel Labs and MIT \\
    \texttt{\small tatbul@csail.mit.edu} \\
    \And
    Vikram Nathan \\
    MIT \\
    \texttt{\small vikramn@mit.edu} \\
    \And
    Vasimuddin Md \\
    Intel Labs \\
    \texttt{\small vasimuddin.md@intel.com} \\
    \And
    Tim Kraska \\
    MIT \\
    \texttt{\small kraska@mit.edu} \\
}

\begin{document}

\maketitle

\begin{abstract}
  Next-generation sequencing (NGS) technologies have enabled affordable sequencing of billions of short DNA fragments at high throughput, paving the way for population-scale genomics.
  Genomics data analytics at this scale requires overcoming performance bottlenecks, such as searching for short DNA sequences over long reference sequences.
  In this paper, we introduce \GENIE (Learned Indexes for Sequence Analysis), a novel learning-based approach to DNA sequence search.
  As a first proof of concept, we focus on accelerating one of the most essential flavors of the problem, called \emph{exact search}. LISA builds on and extends FM-index, which is the state-of-the-art technique widely deployed in genomics toolchains. Initial experiments with human genome datasets indicate that \GENIE achieves up to a factor of $4\times$ performance speedup against its traditional counterpart.
\end{abstract}

\section{Introduction} \label{sec:intro}

Rapid advances in high-throughput next-generation sequencing (NGS) technologies have enabled affordable sequencing of billions of short DNA fragments (called ``reads'') at an unprecedented scale. For example, the Illumina NovaSeq 6000 Sequencer can sequence $20$ billion reads of length $150$ each in less than $2$ days, generating $6$ Terabases of data at a low cost of about \$1000 per human genome \cite{novaseq-6000}. Already today, a growing number of public and private sequencing centers with hundreds of NGS deployments pave the way for population-level genomics. However, realizing this vision in practice heavily relies on building scalable systems for high-speed genomics data analysis.

DNA sequence alignment plays a critical role in genome analysis. In its simplest form, an aligner tries to piece together the short reads by mapping each individual read to a long reference genome (e.g., the human genome consisting of $3$ billion bases). The key operation that has been shown to constitute a significant performance bottleneck during this mapping process is the search for {\em exact} or {\em inexact} matches of read substrings over the given reference sequence \cite{vasim-bb-bioaxiv-2018, langmead2009ultrafast, langmead2012fast, li2009soap2, luo2013soap3, li2009fast}. 
In this paper, we focus on the {\em exact search} variant of this search problem.

The state-of-the-art technique to perform exact search is based on building an FM-index over the reference genome ~\cite{fmindex}.
The key idea behind an FM-index is that, in the lexicographically sorted order of all suffixes of the reference sequence, all matches of a short DNA sequence (a.k.a., a ``query'') will fall in a single region matching the prefixes of contiguously located suffixes. 
Over the years, many improvements have been made to make the FM-index more efficient, leading to several state-of-the-art implementations that are highly cache- and processor-optimized \cite{langmead2009ultrafast, langmead2012fast, li2009soap2, luo2013soap3, li2009fast, nstep-chacon-2013, bwt-chacon-2015, nvbio, zhang2013optimizing, fernandez-bwt-2011, grabowski-bwt-2017, misra2018, bwt-chacon-2015, misra2018}.
Hence, it becomes increasingly more challenging to further improve this critical step in the genomics pipeline to scale with increasing data growth. 

In this paper, we propose a different approach to improving the sequence search performance: LISA (Learned Indexes for Sequence Analysis).
The core idea behind LISA, which enables a new ML-enhanced algorithm for DNA search, is to speed up the process of finding the right region of suffixes by learning the distribution of suffixes in the reference. We do this in a way similar to how learned indexes capture value distributions through models learned from data \cite{learnedindexes}.

When evaluated on an \partCPUtm i9-9900K 3.6 GHz processor, despite being single-threaded and not yet fully optimized to the underlying hardware architecture, our current implementation achieves nearly $4\times$ speedup against a state-of-the-art single-threaded, CPU-optimized version of the FM-index based algorithm \cite{misra2018}, for a workload of $50$ million queries matched against the human genome. This early result shows that learned DNA sequence search is a promising idea
\footnote{Intel Xeon and Intel Xeon Phi are trademarks of Intel Corporation or its subsidiaries in the U.S. and/or other countries. Other names and brands may be claimed as the property of others. \copyright Intel Corporation}.

To the best of our knowledge, this is the first work exploring how ML-enhanced algorithms can improve the process of DNA sequence search, while providing identical semantic guarantees as the traditional algorithms.
\edit{This work is a preliminary proof of concept that can be used as a building block towards fully-optimized learning-based tools for DNA sequence search.}
In summary, this paper makes the following contributions: 

\begin{itemize}
[nosep,leftmargin=1em,labelwidth=*,align=left]
    \item enhancements to the FM-index that enable the application of learning-based search,
    \item a new search algorithm that applies the learning-based approach to the enhanced FM-index to find all exact matches,
    \item an experimental comparison of our approach against a highly-tuned baseline using realistic workloads on the human genome.
\end{itemize}

In the rest of this paper, we first provide some brief background on the traditional FM-index based exact search algorithm as well as the idea of learned index structures which inspired this work; then we present our new approach LISA, along with results from our experimental study.

\section{Background} \label{sec:background}
A DNA sequence is a string over the alphabet, $\Sigma = $ \{A, C, G, T\}, representing the four bases. For the rest of this paper, we use the terms ``sequence'' and ``string'' interchangeably, as well as the terms ``base'' and ``character''. Exact DNA sequence search is a key kernel in many genomics tools, including the widely-used sequence mapping tool Bowtie 2~\cite{langmead2012fast}. Given a reference sequence $R$ and a query sequence $Q$, the goal of exact sequence search is to find exact end-to-end matches of $Q$ in $R$. Typically, $|R| \approx 10^9$ bases; e.g., the length of the human genome is around 3 billion bases. On the other hand, $|Q|$ is typically less than 200 bases; e.g., the default query length in Bowtie 2 is 21.

\begin{figure}[htp]
    \centering
    \includegraphics[width=11cm]{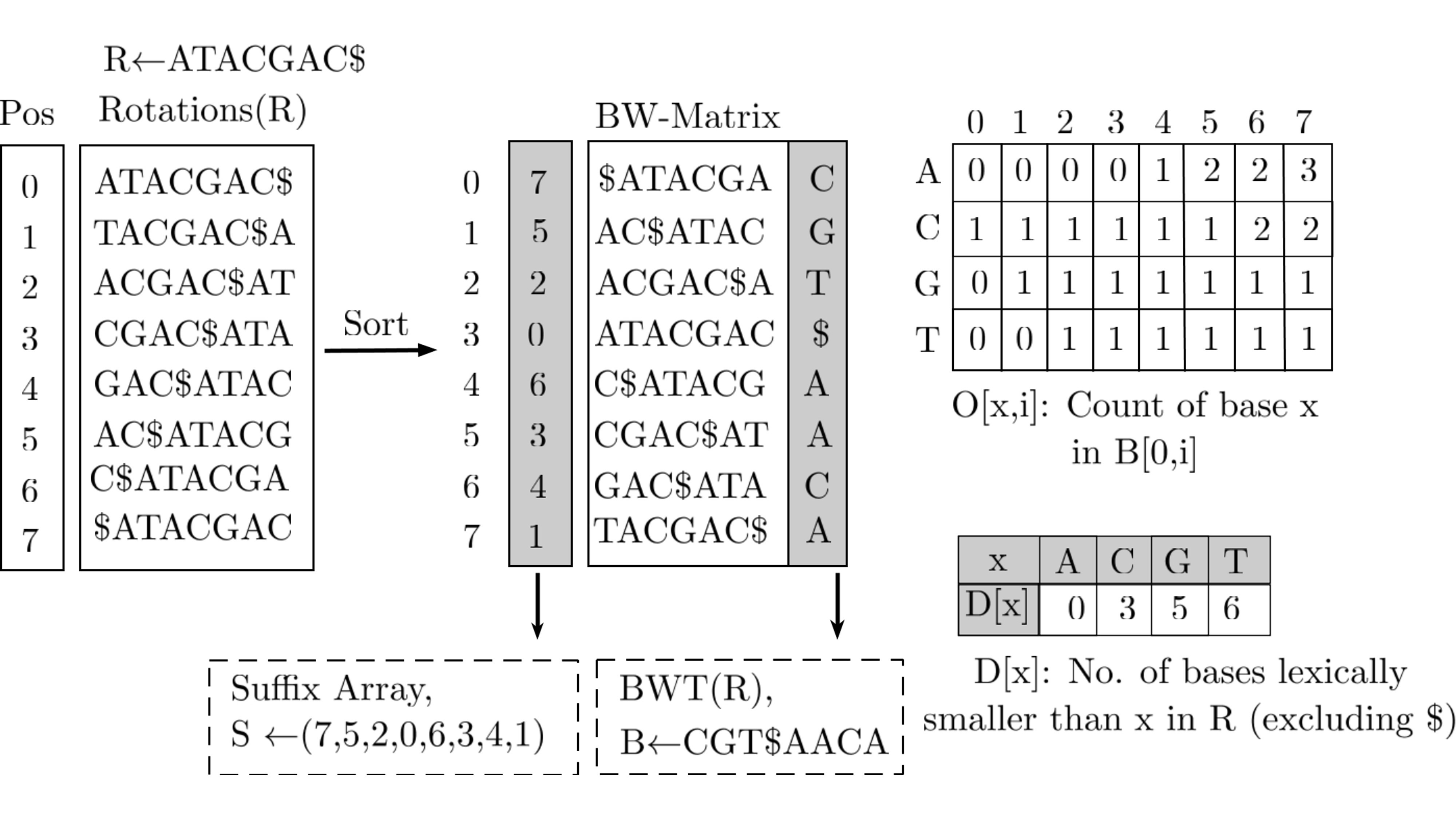}
    \caption{FM-index ($S$, $B$, $D$, $O$) and BW-matrix for sample reference sequence $R\leftarrow$\texttt{ATACGAC\$}. The lexicographical ordering is \$$<$A$<$C$<$G$<$T~\cite{bwa-mem2}.}
    \label{fig:fm-index}
\end{figure}

\noindent
{\bf{FM-index:}} \Figure{fm-index} depicts the construction of the FM-index for an example reference sequence $R$. First, we append $R$ with the character $\$ \notin \Sigma$ which is lexicographically smaller than all characters in $\Sigma$. Subsequently, we obtain all the rotations of $R$ (Rotations(R)). The lexicographically sorted order of the rotations forms the BW-matrix. The BWT ($B$) is the last column of the BW-matrix. The original positions in $R$ of the first bases of these rotations constitute the suffix array ($S$). 

All the exact matches of a query can be found as prefixes of the rotations in the BW-matrix. Since the BW-matrix is lexicographically sorted, these matches are located in contiguous rows of the BW-matrix. Therefore, for a query, all the matches can be represented as a range of rows of the BW-matrix. This range is called the \emph{SA interval} of the query. For example, in \Figure{fm-index}, the \emph{SA interval} of query \emph{``AC''} is $[1,2]$. The values of the suffix array in the SA interval are 5 and 2. Indeed, the sequence \emph{``AC''} is found at positions 5 and 2 in the reference sequence.

The FM-index is used to expedite search for the SA interval~\cite{ferragina2001experimental}. It consists of the suffix array $S$ and the BWT $B$, as well as $D$ and \textit{O} data structures. $D(x)$ is the count of bases in $R[0,|R|-1]$ that are lexicographically smaller than $x \in \Sigma$. $O(x,i)$ is the count of occurrences of base $x$ in $B[0,i]$. Note that the BW-matrix is not stored.

Using the FM-index, finding the SA interval is done using the \textit{backward-search algorithm}. For a reference sequence of length $n$, the algorithm initializes the SA interval to $[0,n)$. The SA interval is updated over the course of the algorithm. Specifically, the algorithm processes the query sequence backwards, starting from the last character, prepending one character at a time and updating the SA interval after each prepended character in $O(1)$ time. To update the SA interval, the FM-index supports a function $f: (char, int) \rightarrow int$ that takes the prepended character $c$ and an integer location $i$, and finds in $O(1)$ time the lower-bound location $i’$ in the BW-matrix of string concat$(c, \text{BW-matrix}[i])$. (The lower bound of a string $s$ is the first entry in the BW-matrix which does not compare less than $s$.) Given an SA interval $[l, u)$, after prepending the character $c$, the SA interval is updated to $[f(c,l), f(c,u))$. Since prepending each character takes $O(1)$ time, the overall algorithm takes $O(|Q|)$ time. For further information on the FM-index, see~\cite{fmindex}.

\noindent
{\bf{Learned Indexes:}} Recent work on learned index structures has introduced the idea that indexes are essentially models that map inputs to positions and, therefore, can be replaced by other types of models, such as machine learning models ~\cite{learnedindexes}. For example, a B-tree index maps a given key to the position of that key in a sorted array. Kraska et al. show that using knowledge of the distribution of keys can produce a learned model, called the recursive model index (RMI), that outperforms B-trees in query time and memory footprint.

Taking a similar perspective, the FM-index can be seen as a model that maps a given query sequence to the SA interval for that query sequence. Based on this insight, in \GENIE, we use knowledge of the distribution of subsequences within the reference sequence to create a learned index structure that enables faster exact search queries.

\section{\GENIE} \label{sec:genie}

\textit{Backward-search algorithm} performs exact search of a query $Q$ using the FM-index by iterating through the query sequence in backwards order, one base at a time, thereby consuming $O(|Q|)$ time. The key idea of \GENIE is to iterate backwards through the query sequence in chunks of $K$ bases at a time, so that exact search takes $O(|Q|/K)$ time. To do this, \GENIE requires two components: (1) a new data structure called the IP-BWT that enables processing $K$ bases at a time, and (2) a method to efficiently search through the IP-BWT, for which we use an RMI. Similar to the \textit{backward-search algorithm}, the goal of \GENIE is to output the SA interval of a query sequence.

\noindent
{\bf{IP-BWT:}} In the \textit{backward-search algorithm}, in order to update the SA interval after prepending a character, FM-index supports a function $f: (char, int) \rightarrow int$ that takes the character $c$ and an integer location $i$, and finds in $O(1)$ time the lower-bound location $i’$ in the BW-matrix of string concat$(c, \text{BW-matrix}[i])$. For \GENIE, we need to support a similar function that takes in a length-$K$ string $s$ and a location $i$, and returns another location $i’$ (i.e. $f_K: (char^K, int) \rightarrow int$) representing the lower-bound location in the BW-matrix of string concat$(s, \text{BW-matrix}[i])$.

\begin{wrapfigure}[19]{r}{0.35\textwidth}
  \centering
    \vspace{-0.5cm}
    \includegraphics[width=0.35\textwidth]{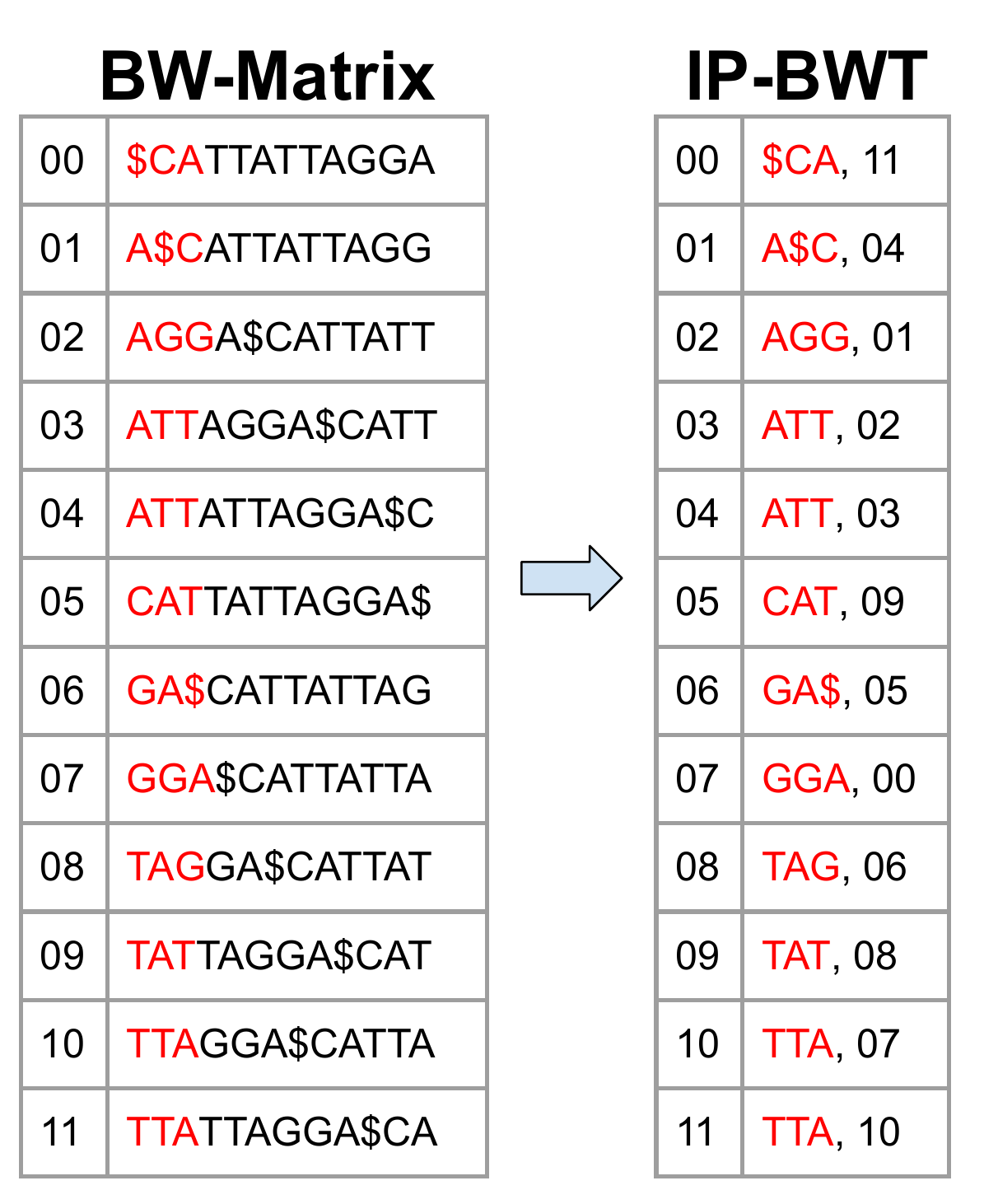}
  \caption{Conversion of a BW-Matrix to IP-BWT on reference \texttt{CATTATTAGGA}, where $K=3$.}
  \label{fig:ip-bwt}
\end{wrapfigure}

For this purpose, we introduce the Index-Paired BWT (IP-BWT) array. 
Each entry of IP-BWT consists of a $(char^K, int)$ pair. The first part is the first $K$ characters of the corresponding BW-matrix row. The second part is the BW-matrix location of the string with the first $K$ and the last $n-K$ characters swapped. Our desired function $f_K: (char^K, int) \rightarrow int$ is now equivalent to finding the lower-bound location of the input $(char^K, int)$ pair in the IP-BWT array. We are free to choose any implementation for how to find that lower-bound location; for example, since the IP-BWT is sorted, we could do a binary search over the entries of the IP-BWT. \Figure{ip-bwt} shows how to create an IP-BWT with $K=3$.

\begin{algorithm}[t]
\caption{Exact Search Algorithm using IP-BWT}
Input: $Q$, a query string; $f_K: (char^K, int) \rightarrow int$, a function that finds the lower-bound location of $(char^K, int)$ in the IP-BWT

Output: $[low, high)$, an SA interval.
\begin{algorithmic}[1]
    \STATE $low, high \leftarrow 0, n$
    \STATE split $Q$ into $\left\lceil{|Q|/K}\right\rceil$ chunks, each of length $K$, with the final chunk possibly shorter than $K$
    \FOR{$\mathcal{C}$ in reversed order of chunks}
        \IF{$|\mathcal{C}|<K$}
            \STATE \textit{/* Special case for when final chunk has length less than $K$ */}
            \STATE $\mathcal{C}_{low} \leftarrow \mathcal{C} + ``\$" + ``A" \times (K-|\mathcal{C}|-1)$
            \STATE $\mathcal{C}_{high} \leftarrow \mathcal{C} + ``T" \times (K-|\mathcal{C}|)$
            \STATE $low, high \leftarrow f_K((\mathcal{C}_{low}, low)), f_K((\mathcal{C}_{high}, high))$
        \ELSE 
            \STATE $low, high \leftarrow f_K((\mathcal{C}, low)), f_K((\mathcal{C}, high))$
        \ENDIF
    \ENDFOR 
    \RETURN $[low, high)$
\end{algorithmic}
\label{alg:exact_search}
\end{algorithm}

\Algorithm{exact_search} shows the exact search algorithm using IP-BWT. For example, using the reference sequence and IP-BWT from \Figure{ip-bwt}, let the query sequence be \texttt{ATTA}. We split this into two chunks: \texttt{ATT} and \texttt{A} (line 2). We first use the RMI to find the lower bound locations of $(A\$A, 0)$ and $(ATT, n)$, which are 1 and 5, respectively. We then use the RMI to find the lower bound locations of $(ATT, 1)$ and $(ATT, 5)$, which are 3 and 5. Our algorithm gives the interval $[3, 5)$. We can confirm that \texttt{ATTA} can be found in position 3 and 4 of the BW-matrix.

\noindent
{\bf{Faster Chunk Processing using RMI:}} Using the IP-BWT, we are able to process the query sequence in chunks of $K$ bases at a time. However, when processing each chunk, we must evaluate the function $f_K$, which involves a binary search over the IP-BWT. This takes $O(\log{n})$ time, where $n$ is the number of entries in the IP-BWT, which is equivalent to the length of the reference sequence. Therefore, the overall runtime of exact search using IP-BWT and evaluating $f_K$ with binary search is $O(|Q|\log{n}/K)$. For large reference sequences, this might be slower than \textit{backward-search} using the FM-index.

\begin{figure}[htp]
    \centering
    \includegraphics[width=\textwidth]{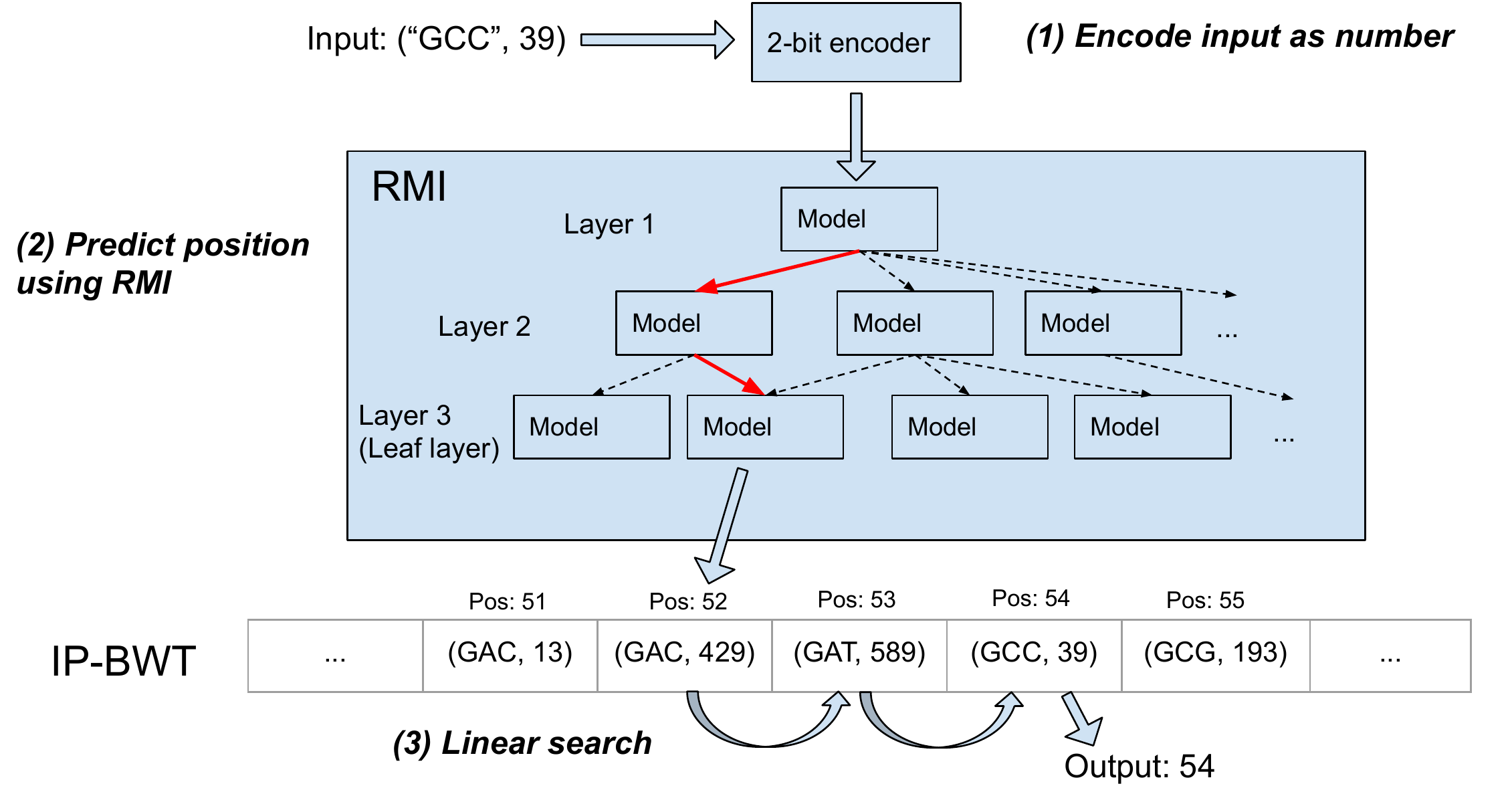}
    \caption{Using a 3-layer RMI to evaluate $f$ for an example input $(GCC, 39)$. The RMI predicts position 52, and linear search finds the correct position. The solid red lines show the traversal path down the RMI.}
\label{fig:rmi}
\end{figure}

In order to avoid paying the cost of a binary search for each evaluation of $f_K$, we use a learned approach to support $O(1)$ evaluation of $f_K$. In particular, $f_K$ is a model that maps input keys ($(char^K,int)$ pairs) to their positions in the sorted IP-BWT. We model $f_K$ using the RMI, which is a hierarchy of models that is quick to evaluate~\cite{learnedindexes}; the RMI conceptually resembles a hierarchical mixture of experts~\cite{mixtureofexperts}. \Figure{rmi} shows an example of using a 3-layer RMI to evaluate $f_K$ in three steps: (1) since the RMI only accepts numbers as inputs, we first convert the input $(char^K,int)$ into a number. Since the alphabet $\Sigma$ only has 4 characters, any character can be represented in 2 bits. Therefore, we convert $char^K$ into a number with $2K$ bits by concatenating the bits of the individual characters together. We then append the bits of the $int$. Note that we have a special case for handling the sentinel character $\$$ while maintaining this 2-bit encoding. (2) We give the encoded input to the RMI and traverse down the layers of the RMI to a leaf model. The leaf model predicts the position in the IP-BWT where it expects to find the input pair. (3) If the predicted position does not contain the input pair, we use linear search over the IP-BWT starting from the predicted position to find the actual position of the pair. \edit{Note that this learning-based approach to modeling $f_K$ guarantees correctness; \GENIE will produce exactly the same results as using backward search with FM-index.}

\edit{Unlike the RMI proposed in~\cite{learnedindexes}, which constructs the model hierarchy top-down according to the user-selected number of models at each layer, we construct the RMI bottom-up according to desired bounds on the average error between the predicted position and the actual position. We use these desired bounds to determine the number of models at each layer. Bounds on the average prediction error are useful for limiting the time spent on step 3 of the RMI evaluation workflow described above, because they directly reflect the average number of iterations of linear search. Given a desired bound $\alpha$ on the average error, we begin building the RMI at the leaf layer by partitioning the IP-BWT entries into contiguous blocks that can each be modeled with average error no more than $\alpha$. We find this partitioning by starting with one partition that comprises of the entire IP-BWT, then recursively splitting equally in two until each partition achieves the $\alpha$ bound. A leaf model is built on the entries of each partition. The smallest entry in each partition is used to repeat this procedure in order to find the partitions in the layer above the leaf layer, and so forth until we have one model at the root layer. Since the non-leaf models are allowed to have prediction error, we may need to perform linear search at each layer of the RMI, instead of only the leaf layer. Note that we can set different values of $\alpha$ for each layer of the RMI. Also, note that at the root layer, there is only one partition; therefore, we do not set an $\alpha$ bound for the root model. Since we have no guarantee on average error for the root model, the root model uses exponential search instead of linear search.}

\noindent
{\bf{Optimization for Batched Queries:}} So far, we have presented our method of processing a single query. However, in practice, queries arrive in large batches. For large query batches, we can adopt an additional optimization. Because the final layer of the RMI (i.e., RMI leaves), which are responsible for predicting locations, are arranged in sorted order, if the queries also arrive in sorted order, we can use the \textit{double-pointer technique} (\Algorithm{double-ptr})
to find for each query its corresponding RMI leaf. This resembles the merge step of merge-sort. When the number of queries is $\Omega($number of RMI leaves$)$, the amortized time for each query to find its RMI leaf is $O(1)$. This avoids the cost of traversing down the RMI for each query, and also improves cache locality.

\edit{Note that this optimization for batches cannot be easily applied to the FM-index. This optimization requires that inputs to $f_K$ are sorted; since IP-BWT processes in chunks of $K$ characters, we only need to sort queries $O(|Q|/K)$ times. However, since backward search with the FM-index prepends only one character at a time, it would need to sort the inputs to $f$ after each prepended character. Sorting inputs $O(|Q|)$ times would impose a significant performance bottleneck, which makes this optimization impractical when using FM-index.}

\begin{algorithm}[H]
\caption{The RMI-based exact search algorithm for large query batch}
Input: $qs[]$, a list of query strings encoded as numbers.

Output: a list of SA intervals.
\begin{algorithmic}[1]
    \STATE sort $qs$
    \STATE $leafPtr \leftarrow \&rmi.leaves[0]$
    \FOR{$q$ in $qs$}
        \WHILE{$q \geq leafPtr.ipbwt\_range\_upper\_bound$}
            \STATE increment $leafPtr$
        \ENDWHILE 
        \STATE $prediction \leftarrow leafPtr.predict(q)$
        \STATE perform linear search around $prediction$ to find the SA interval of $q$
    \ENDFOR 
\end{algorithmic}
\label{alg:double-ptr}
\end{algorithm}

\noindent
{\bf{Discussion:}} \GENIE's speed advantage over FM-index comes from two components: (1) the IP-BWT, which allows \GENIE to process $K$-character chunks of the query at a time, whereas FM-index processes one character at a time, and (2) using an RMI to process each $K$-character chunk in $O(1)$ time, whereas a naive binary search would take $O(\log{n})$ time per chunk. Therefore, an exact search query using FM-index takes $O(|Q|)$ time, whereas \GENIE takes $O(|Q|/K)$ time.

To discuss memory consumption, we measure in bytes in terms of $n$, the length of the reference sequence. The reference sequence itself therefore takes $0.25n$ space. In the state-of-the-art implementation of \textit{backward-search algorithm}, the space of suffix array is $4n$, a compressed structure that combines BWT and $O$ is $2n$, and $D$ is negligible, for a total space consumption of $6n$. For \GENIE, the space of the suffix array is $4n$, IP-BWT is $(0.25K+4)n$, and the RMI is usually around $0.5n$, for a total space consumption of $8.5n+0.25Kn$. For example, in Bowtie 2 the default value of $|Q|$ is 21, so using an IP-BWT with $K=21$, \GENIE takes around $13.75n$ space, which is larger than the FM-index. However, slightly larger space consumption is usually not a concern in practice, and if necessary \GENIE can use smaller $K$ or compress the IP-BWT. Though the space usage of \GENIE would increase with larger values of $K$, we find through experiments that \GENIE maintains good performance using an IP-BWT with $K=21$, even for large query lengths. Therefore, the space of \GENIE does not need to grow beyond around $13.75n$.

Could we replace the RMI with some other index structure that can evaluate $f_K$ over the IP-BWT even faster? \cite{learnedindexes} showed that RMIs perform better than binary search and B trees. A lookup table (implemented as an array) that stores the output of $f_K$ for every possible pair $(char^K,int)$ would also allow $O(1)$ evaluations but would far exceed memory capacity, \edit{even if we use an IP-BWT with very small $K$}. However, it is possible to combine a downsampled version of the lookup table with binary search, which we discuss in the evaluation. Another idea is to use a hash table to map all $n$ existing pairs $(char^K,int)$ to their positions; \edit{however, this fails because we almost always need to evaluate $f_K$ on pairs that do not exist. For example, in the example attached to \Figure{ip-bwt}, we find the lower bounds of four pairs---$(A\$A, 0)$, $(ATT, n)$, $(ATT, 1)$, and $(ATT, 5)$---none of which exist in the IP-BWT}. Also, note that since the entries of the IP-BWT must be sorted in order to maintain one contiguous SA interval, the IP-BWT itself cannot be replaced with a hash table to enable faster searches.

\section{Evaluation} \label{sec:eval}
We present preliminary results\footnote{Software and workloads used in performance tests may have been optimized for performance only on Intel microprocessors. Performance tests, such as SYSmark and MobileMark, are measured using specific computer systems, components, software, operations and functions. Any change to any of those factors may cause the results to vary. You should consult other information and performance tests to assist you in fully evaluating your contemplated purchases, including the performance of that product when combined with other products. For more information go to www.intel.com/benchmarks.

Benchmark results were obtained prior to implementation of recent software patches and firmware updates intended to address exploits referred to as "Spectre" and "Meltdown".  Implementation of these updates may make these results inapplicable to your device or system.}
for \GENIE. Experiments use SIMD, running a single-thread implementation on an Ubuntu system with \partCPUtm i9-9900K 3.6GHz processor and 64GB RAM. As our baseline, we use a highly CPU-optimized implementation of backward search algorithm that is significantly faster than its alternatives~\cite{misra2018}. We also compare with doing backwards search using IP-BWT and binary search (i.e., without the RMI).

\noindent
{\bf{Workload Scenario:}} 
We evaluate \GENIE on a real-world scenario: we use whole human genome as the reference sequence and for query sets, we use large sets of short query sequences of various lengths as would be found in several prominent sequence aligners. We train \GENIE using an IP-BWT with $K=21$. The RMI has three layers, only uses linear regression models, and \edit{we construct the RMI using $\alpha=14$ for the second layer models and $\alpha=6$ for the leaf layer models. $\alpha$ is tuned once for optimal performance on the reference sequence;}
the RMI is not re-trained or re-tuned for each experiment. We evaluate on four different query sequence lengths: (1) $|Q|=21$, so that \GENIE processes exactly one chunk, (2) $|Q|=42$, so that \GENIE processes multiple whole chunks, (3) $|Q|=32$, so that \GENIE processes a chunk shorter than $K$, and (4) a very long query, $|Q|=200$.
For each of these query sequence lengths, we generate a batch of 50 million query sequences randomly from the human genome to be our query set, and run them together with \GENIE, using the batched-query optimization from \Section{genie}.

\Table{experiment1} shows that \GENIE is 2.73$\times$ to 3.97$\times$ faster than the optimized FM-index baseline on these query lengths. Query lengths that are a perfect multiple of $K$ (i.e., 21 and 42) perform the best. The query length of 32 has slightly lower relative performance, as \GENIE must still process 2 chunks per query, as if the query has length 42. The very long query sequence has lower relative performance, as \GENIE's performance does not scale linearly: for longer queries, the time spent on sorting grows super-linearly. \GENIE achieves around a 2$\times$ performance boost from using the RMI instead of binary search.

\begin{table}[htp]
\begin{center}
\begin{tabular}{lcccc}
\toprule
 & $|Q|=21$ & $|Q|=32$ & $|Q|=42$ & $|Q|=200$ \\
 \midrule
Optimized FM-index & 1509   & 2414  & 3284 & 15411  \\
IP-BWT with binary search   & 785 (1.92$\times$)   & 1424 (1.70$\times$)   & 1779 (1.85$\times$) & 10414 (1.48$\times$) \\
\GENIE                      & 383 (3.94$\times$)   & 762 (3.17$\times$)   & 827 (3.97$\times$) & 5646 (2.73$\times$) \\
\bottomrule
\end{tabular}
\end{center}
\caption{Average query time (measured in ticks per query) of exact search on 50 million queries, while varying query length. Relative speedup to FM-index is shown in parentheses.}
\label{tab:experiment1}
\end{table}

\begin{wrapfigure}[14]{r}{0.6\textwidth}
  \centering
    \vspace{-0.2cm}
    \includegraphics[width=0.6\textwidth, clip]{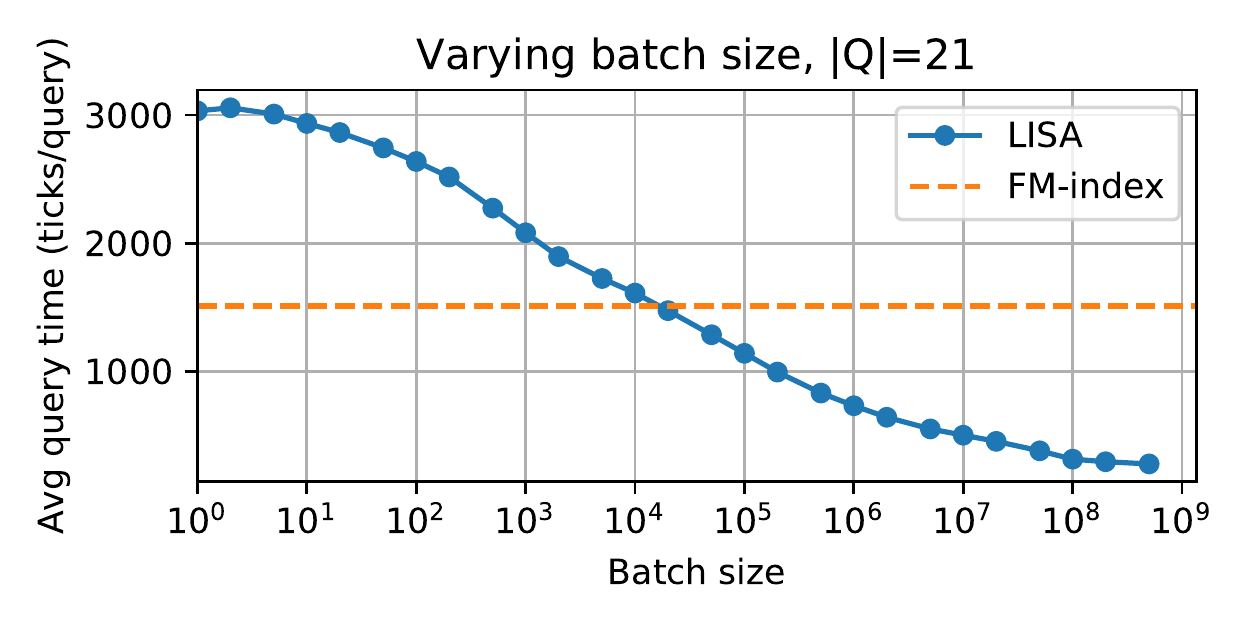}
  \caption{\GENIE's performance advantage increases with batch size.}
  \label{fig:experiment2}
\end{wrapfigure}

\noindent
{\bf{Batch Size:}} In order to measure the effect of batch size on query times, we fix the query sequence length to $|Q|=21$ and vary the batch size from 1 to 500 million. \Figure{experiment2} shows that as batch size increases, \GENIE's time per query decreases. For batches larger than 10 thousand, \GENIE starts to outperform the FM-index.

\GENIE's performance is poor for small batch sizes, where the double pointer technique actually hurts performance, as the merge step will skip over many leaves. However, in real-world scenarios, it is rare to have such small batch sizes where this would matter. \edit{Next-generation sequencing technologies produce billions of DNA fragments, and downstream applications such as DNA sequence alignment need to process all or most of the produced fragments at once. Therefore, the typical use case for \GENIE is to process large batches of queries.} If high performance on small batches is absolutely necessary, it is possible to achieve comparable/better performance than FM-index for small batch sizes by not using the double-pointer technique. 

\noindent
{\bf{Alternative to RMI:}} To evaluate the effectiveness of RMI against alternative index structures, we compared \GENIE to a version which evaluates $f_K$ using a combination of a lookup-table and binary search, instead of using the RMI. If we treat a $(char^K, int)$ pair as a $(2K+32)$-bit number, we maintain a ``downsampled'' lookup-table (implemented as an array) with $2^p$ entries, where $p < 2K+32$. Let $q = 2K+32-p$. The $i$-th entry of the lookup-table holds the lower-bound location in the IP-BWT of the pair represented by the number $i\cdot q$. Essentially, our downsampled lookup-table holds every $q$-th entry of a ``full'' lookup-table that contains every possible $(char^K, int)$ pair. To evaluate $f_K$ on an input represented as a $2K+32$-bit number, we do a lookup in the lookup-table using the first $p$ bits to obtain lower and upper bounds on the location of the input pair in the IP-BWT, then do a binary search with the remaining $q$ bits on the IP-BWT starting from those bounds to find the location of the input. This is our best effort at an alternative index structure that is most competitive with RMI.

For $K=21$, $|Q|=21$, and batch size of 50 million, replacing the RMI with a downsampled lookup-table of the same memory size results in 40\% slower query times. More performant lookup-tables use significantly more memory: the lookup-table itself has space $2^{p+2}$ bytes, so even with only $p=34$, the lookup table takes 69GB, which exceeds the memory of our machine. Therefore, the lookup-table is impractical and we do not include it in this paper.

\section{Conclusions} \label{sec:conclusion}
In this work, we introduced \edit{a preliminary version of} our learned indexing approach for DNA sequence search, \GENIE, which produces promising \edit{initial} results for the exact search problem when tested on workloads of realistic queries against the human genome. In particular, \GENIE achieves up to nearly $4\times$ faster query times than an extensively optimized version of the FM-index based approach, which has been the common practice in sequence search. We believe that the core ideas behind \GENIE can be extended to other types of DNA sequence search problems. In particular, we are currently working on using learned indexes to speed up searching for super maximal exact matches (SMEMs) for a query in the reference. For any position in query, an SMEM is the longest substring of the query through that position that has an exact match in the reference \cite{bwa-mem2, li2012, bwamem}. We are also working with the Broad Institute to integrate \GENIE into applications that are widely used by the genomics community.

\subsubsection*{Acknowledgments}

We thank Tony Peng, Ashwath Thirumalai, and Elizabeth Wei for their contributions to the original design of \GENIE; and Pradeep Dubey and Heng Li for their valuable feedback. This research has been funded in part by affiliate members and supporters of DSAIL (Data Systems and AI Lab) at MIT -- Google, Intel, and Microsoft.

\bibliographystyle{unsrtnat}
\bibliography{genie}

\end{document}